\documentclass[fleqn,10pt]{SelfArx}  
\usepackage{setspace}
\usepackage[]{algorithm}
\usepackage{natbib}
\usepackage{makecell}
\newcommand \bcol{}

\usepackage{pdflscape}

\newcommand\blfootnote[1]{%
  \begingroup
  \renewcommand\thefootnote{}\footnote{#1}%
  \addtocounter{footnote}{-1}%
  \endgroup
}

\definecolor{color1}{RGB}{0,0,0} 
\definecolor{color2}{RGB}{0,20,20} 
\definecolor{color3}{RGB}{0,0,140} 
\definecolor{color4}{RGB}{0,50,200} 

\usepackage{journalabbr}

\usepackage{hyperref} 
\hypersetup{hidelinks,colorlinks,breaklinks=true,urlcolor=color3,citecolor=color3,linkcolor=color4,bookmarksopen=false,pdftitle={Title},pdfauthor={Author}}

\JournalInfo{Accepted for publication in JATIS} 
\Archive{arXiv preprint typeset using template Stylish-Article$^{+}$} 

\PaperTitle{Gamma-ray burst localisation strategies for the SPHiNX hard
X-ray polarimeter} 

\Authors{L. Heckmann\textsuperscript{1}, N.~K. Iyer\textsuperscript{2,3}*,
M. Kiss\textsuperscript{2,3}, M. Pearce\textsuperscript{2,3}, F.
Xie\textsuperscript{2,3}$^{\dagger}$ } 
\affiliation{\textsuperscript{1}\textit{Technische Universit\"at Wien, Faculty of Physics, 1040
Vienna, Austria}} 
\affiliation{\textsuperscript{2}\textit{KTH Royal Institute of Technology, Department of Physics, 106 91
Stockholm, Sweden}} 
\affiliation{\textsuperscript{3}\textit{The Oskar Klein Centre for Cosmoparticle Physics, AlbaNova University
Centre, 106 91 Stockholm, Sweden}} 
\affiliation{$^{\dagger}$ \textit{Now at INAF, Osservatorio Astronomico di Cagliari, via della Scienza 5, 09047
Selargius, Italy}}
\affiliation{*\textbf{Corresponding author}: \href{mailto:nkiyer@kth.se}{nkiyer@kth.se}} 

\newcommand{\keywordname}{Keywords} 

\Abstract{SPHiNX is a proposed gamma-ray burst (GRB) polarimeter mission operating in
	the energy range 50-600 keV with the aim of studying the prompt emission
	phase. The polarisation sensitivity of SPHiNX reduces \bcol{as the}
	uncertainty on the GRB sky position \bcol{increases}.  
	The stand-alone ability of the SPHiNX design to localise GRB positions is
	explored via Geant4 simulations.
	Localisation at the level of a few degrees is possible using three different
	routines. This results in a large fraction ($>$ 80\%) of observed GRBs
	having a negligible ($<5\%$) reduction in polarisation sensitivity due to the
	uncertainty in localisation.}

\begin{document}

\flushbottom 

\maketitle 

{
\hypersetup{linkcolor=black}
\tableofcontents 
}

\thispagestyle{empty} 


\blfootnote{$^{+}$\url{http://www.latextemplates.com/template/stylish-article}}
\section{Introduction}
Gamma-ray bursts (GRB) were serendipitously discovered over 50 years ago by Vela satellites deployed to monitor the ban on nuclear tests in space~\citep{GRB}.
They are now known to be the brightest events in the electromagnetic universe,
occurring approximately once per day at random locations on the
sky~\citep{paciesas1999,fermicat}. GRBs are thought to be formed during the
collapse of a massive object into a black hole~\citep{Meszaros06}. Two highly
relativistic back-to-back plasma jets are produced aligned with the
rotational axis of the black hole. The electromagnetic \bcol{signature} of a jet oriented in the earth-direction is
detected by instruments as a GRB. Such observations reveal that
the emission exhibits a high-energy (keV--MeV) prompt phase (seconds to minutes
duration) caused by energy dissipation within the jet; and, a lower energy
afterglow (days duration) created by interactions between the burst ejecta
and interstellar gas. Both phases have been extensively studied. A
relatively coherent description of the afterglow has emerged, but there are
still fundamental open questions concerning the underlying physical processes
behind the prompt emission~\citep{Zhang&Kumar}.

Measurements of the linear polarisation
of prompt emission can discriminate between these emission models
without the degeneracies associated with modeling of spectral and temporal
observations~\citep{toma09}.
Linear polarisation is described by: $(i)$ the polarisation fraction (PF, \%)
describing the magnitude of beam polarisation; and, $(ii)$ the polarisation
angle (PA, degrees) defining the orientation of the electric field vector.
Reliable measurements require purpose-built and well-calibrated instruments,
and only rudimentary polarimetric observations have been performed to
date~\citep{mcconnell}.
A particular challenge is that the polarimetric response of the instrument
depends on the relative location of the GRB~\citep{muleri}. 
Thus localisation strategies become important for instruments studying GRB polarisation. 
Localisation is also important
for multi-wavelength/messenger studies of GRBs. This is exemplified by the recent
multi-messenger observation campaign for GRB170817A~\citep{abbott17b}. Localising
this burst with an accuracy of a few degrees using \textit{Fermi} GBM~\citep{fermigbm} played an important
role in enabling searches with other telescopes~\citep{abbott17c}.

SPHiNX is a hard X-ray polarimeter (50--600  keV) proposed for the Swedish InnoSat small
satellite platform~\citep{pearce18}. 
The main objective of the SPHiNX mission is to obtain statistically significant polarisation
measurements for a large number of GRBs ($\sim$50) to enable discrimination
between GRB prompt emission models. This is achieved through a
large field of view (120$^\circ$ cone angle), high polarisation sensitivity and two years of operation in orbit.

The GRB localisation performance of SPHiNX is studied in this paper.
The SPHiNX mission and methodology for high-energy polarimetry is described in
Section~\ref{sec:sphinx}. 
In Section~\ref{sec:prin}, the principle used to localise GRBs is outlined. 
Details of the methods used for localisation and obtained results are given in Section~\ref{sec:uncer}. 
The paper concludes with a discussion and conclusions in Sections~\ref{sec:discussion} and~\ref{sec:conclusion}, respectively.

\begin{figure*}[!ht]
	\centering
	\begin{tabular}{l}
	\includegraphics[width=\textwidth]{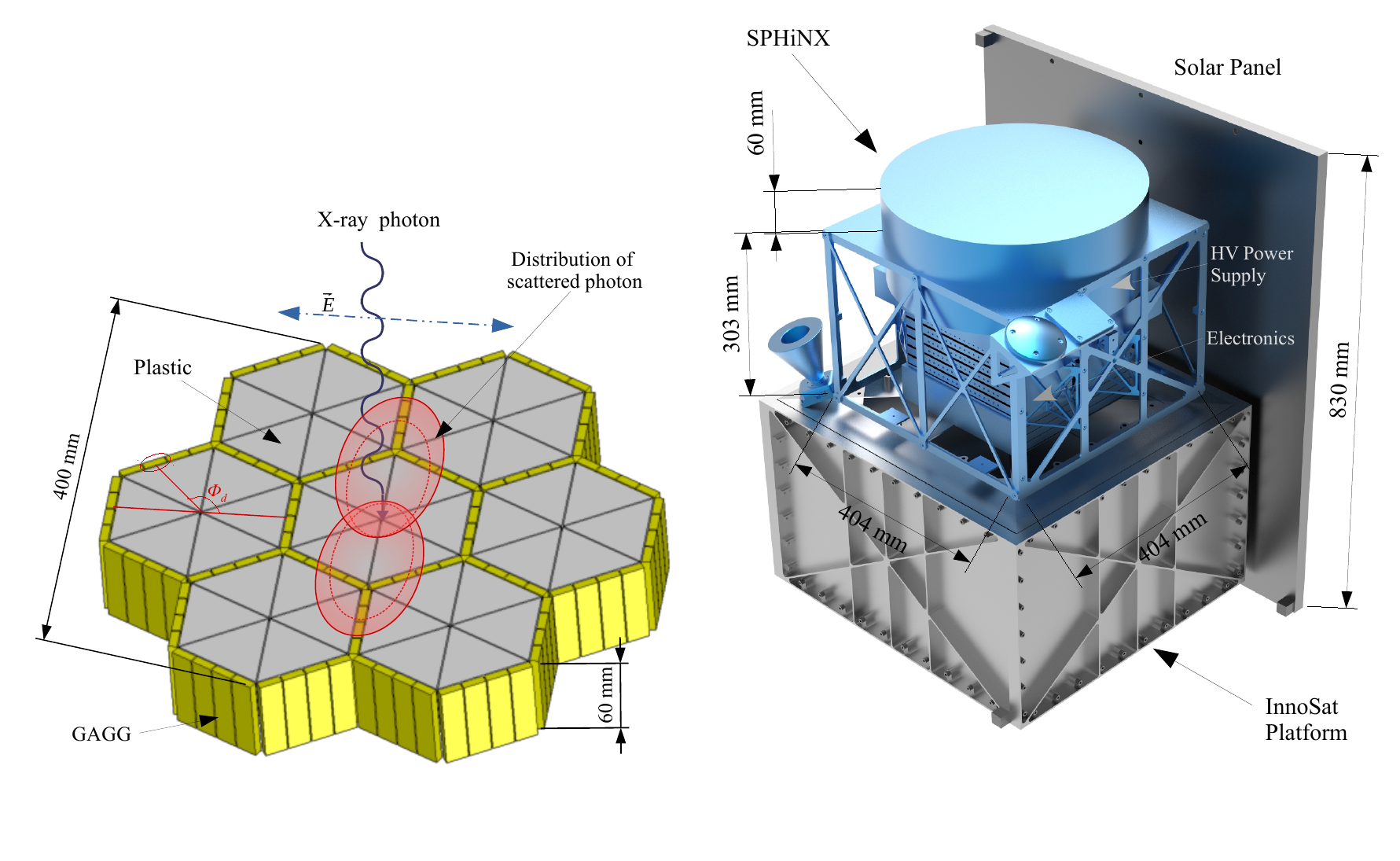}
	\\
	\hspace{3.5cm}(a) \hspace{8.1cm} (b)
	\end{tabular}
	\caption{SPHiNX mission : (a) Azimuthal distribution of the scattered photon
		projected on the SPHiNX scintillator assembly. Detector angle $\phi_d$
		(see Section \ref{ss:resmod}) is also shown. (b) Depiction of the SPHiNX
	payload as assembled on the InnoSat platform.}
	\label{fig:scat}
\end{figure*}

\section{The SPHiNX mission proposal}\label{sec:sphinx}
SPHiNX utilises Compton scattering of incident photons within a scintillator
assembly to
estimate the polarisation properties.
The interaction cross-section for Compton scattering is described by the Klein-Nishina relationship,
\begin{equation}
\frac{d\sigma}{d\Omega} =
\frac{1}{2}
r_{e}^{2}
\frac{k^2}{k_{0}^{2}}
\left(
\frac{k}{k_{0}} +
\frac{k_{0}}{k} -
2\sin^2 \theta\,\cos^2 \phi
\right),
\end{equation} 
where $r_e$ is the classical electron radius, $k_0$ and $k$ are the momenta of
the incoming and scattered photon, and $\theta$ and $\phi$ are polar and azimuthal
scattering angles defined relative to the co-ordinate axes made with the direction
and plane of polarisation of the incident photon. 
X-rays will preferentially scatter in a direction perpendicular to the
polarisation vector, as depicted in Fig.~{\ref{fig:scat}(a)}. The polarisation of
hard X-rays can therefore be determined in a segmented detector by determining
the angle through which the Compton scattering occurs.

The SPHiNX polarimeter comprises 162 detector units (42 plastic and 120 GAGG 
scintillator units), arranged as shown in Fig.~{\ref{fig:scat}(a)}. The periphery of the scintillators is 
covered by a contour hugging multi-layered metal shield (of 60 mm height) to reduce
the background counting rate. 
The polarimeter design is optimised to have high polarisation sensitivity whilst
satisfying the InnoSat mission constraints, resulting in a flat pixelated geometry
(Fig.~{\ref{fig:scat}}). 

The polarisation parameters of incident photons are determined by identifying
coincident Compton scattering (preferentially occurring in the low atomic number
plastic) and photoelectric absorption interactions (preferentially occurring in
the GAGG). Such a combination of scintillator
interactions results in double-hit events and defines the azimuthal scattering
angle $\phi$.
The distribution of $\phi$ is a harmonic function referred to as a modulation
curve, where the phase defines PA. PF is defined as
$M$/$M_{100}$,
where $M$ is the measured modulation amplitude and $M_{100}$ is the modulation amplitude for a 100\% polarised beam. 

\begin{figure}[!h]
	\centering
	\includegraphics[width=0.5\textwidth]{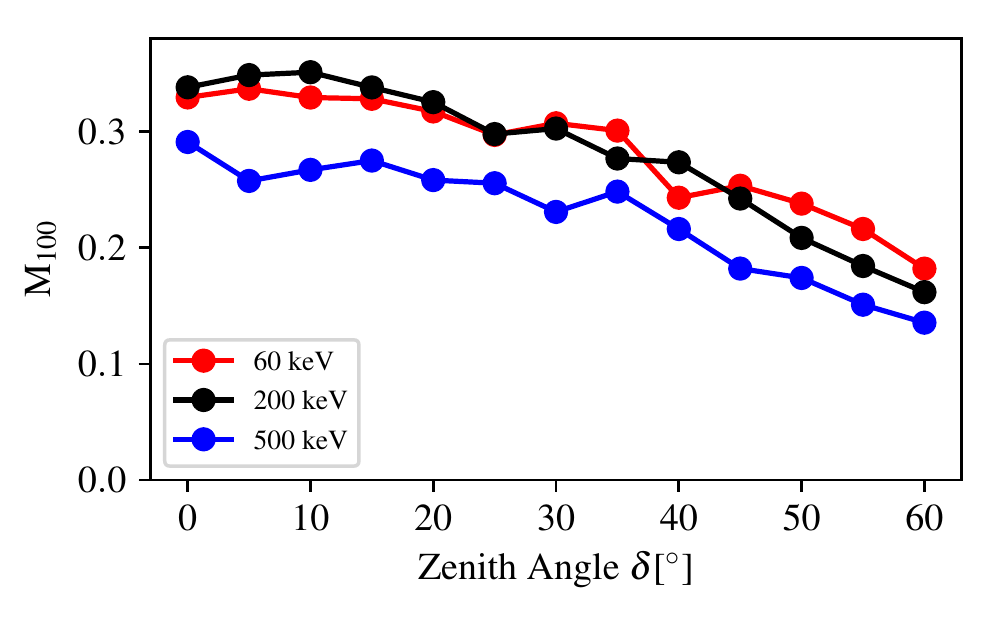}
	\caption{Variation in $M_{100}$ for the SPHiNX detector with incoming photon
		\bcol{zenith} angle. Values are obtained from simulations described in
	\citet{pearce18}. }
	\label{fig:phim100}
\end{figure}

It has become standard to express the polarimetric sensitivity at a 99\%
($\sim$3$\sigma$) confidence level in terms of the Minimum Detectable
Polarisation (MDP)~\citep{MDP}, 
\begin{equation}
\mathrm{MDP} = \frac{4.29}{M_{100}\,R_s} \sqrt{\frac{R_s + R_b}{T}},
\label{eqn:mdp}
\end{equation} 
where $R_s$ ($R_b$) is the signal (background) rate (Hz) and $T$ is the duration of the burst observation (s).
There is a 1\% probability for an unpolarised GRB to yield PF $>$ MDP through statistical fluctuations. 
For a given GRB, the value of $M_{100}$ will depend on the location of the GRB
with respect to the polarimeter since the modulation pattern depends on the
direction from which the detector array is illuminated. An on-axis GRB will
produce a sinusoidal modulation curve, whereas distortions will be introduced
and the modulation amplitude will decrease as the GRB moves off-axis.
This effect is illustrated in Fig.~\ref{fig:phim100} for SPHiNX. 

\begin{figure*}[!h]
	\centering
	\includegraphics[width=0.8\textwidth]{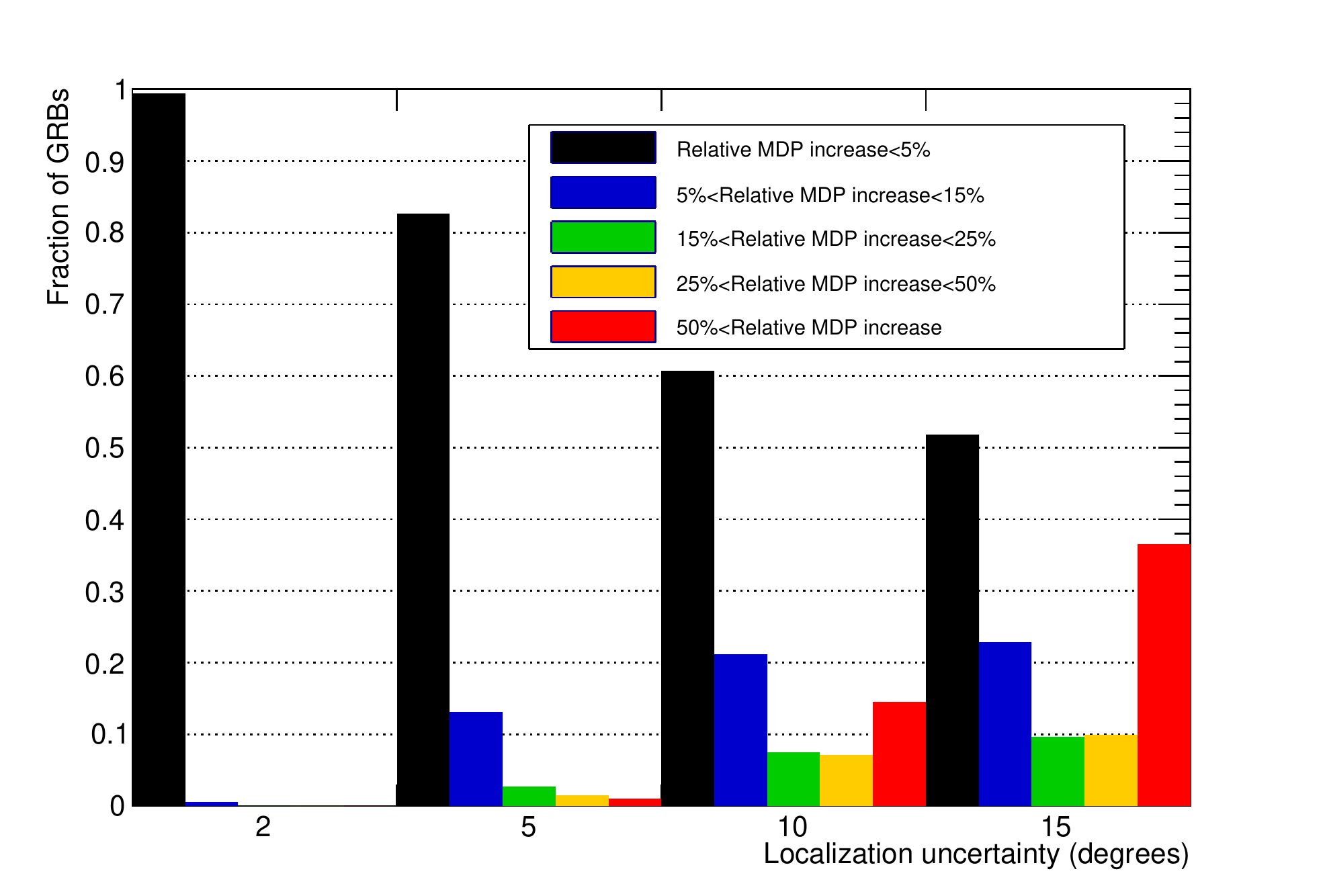}
	\caption{Effect of uncertainty in source position \bcol{(zenith angle)} on the MDP.  Results
		\bcol{are} obtained by Monte-Carlo sampling of GRBs uniformly distributed in SPHiNX
field of view.}
	\label{fig:vic}
\end{figure*}

Once the GRB location is known relative to SPHiNX, the value of $M_{100}$ is
determined from computer simulations validated at discrete
energies by laboratory measurements~\citep{validate}.  
Any uncertainty in the GRB localisation will propagate to an uncertainty in
\bcol{PF,PA} and 
increase the MDP (reduce sensitivity). This is the key
difference between GRB polarimeters and on-axis pointed polarimeters for which
the source direction is known and fixed for all measurements.
Uncertainty in localising these GRBs will also lower the number of GRBs for which significant
polarisation measurements can be obtained (due to \bcol{an} increase in MDP). This can
diminish the scientific potential for SPHiNX as is seen from
Fig.~\ref{fig:vic}. 

Some of the GRBs observed by SPHiNX will be simultaneously observed by other
missions. In this case, the GRB sky position can be determined through 
the Gamma-ray \bcol{Coordination} Network (GCN) and Interplanetary Network (IPN). 
In this paper, we explore the stand-alone accuracy
with which SPHiNX can localise GRBs.
The InnoSat mission parameters (e.g. downlink cadence, on-board data processing
and storage limits) mean that the localisation would be performed on-ground when data
is downloaded once per day. The localisation methods discussed in this paper
are, however, sufficiently generic and may be implemented on-board in real-time
for missions without such constraints. 

\begin{figure*}[!ht]
	\centering
	\begin{tabular}{l}
	\includegraphics[width=0.97\textwidth]{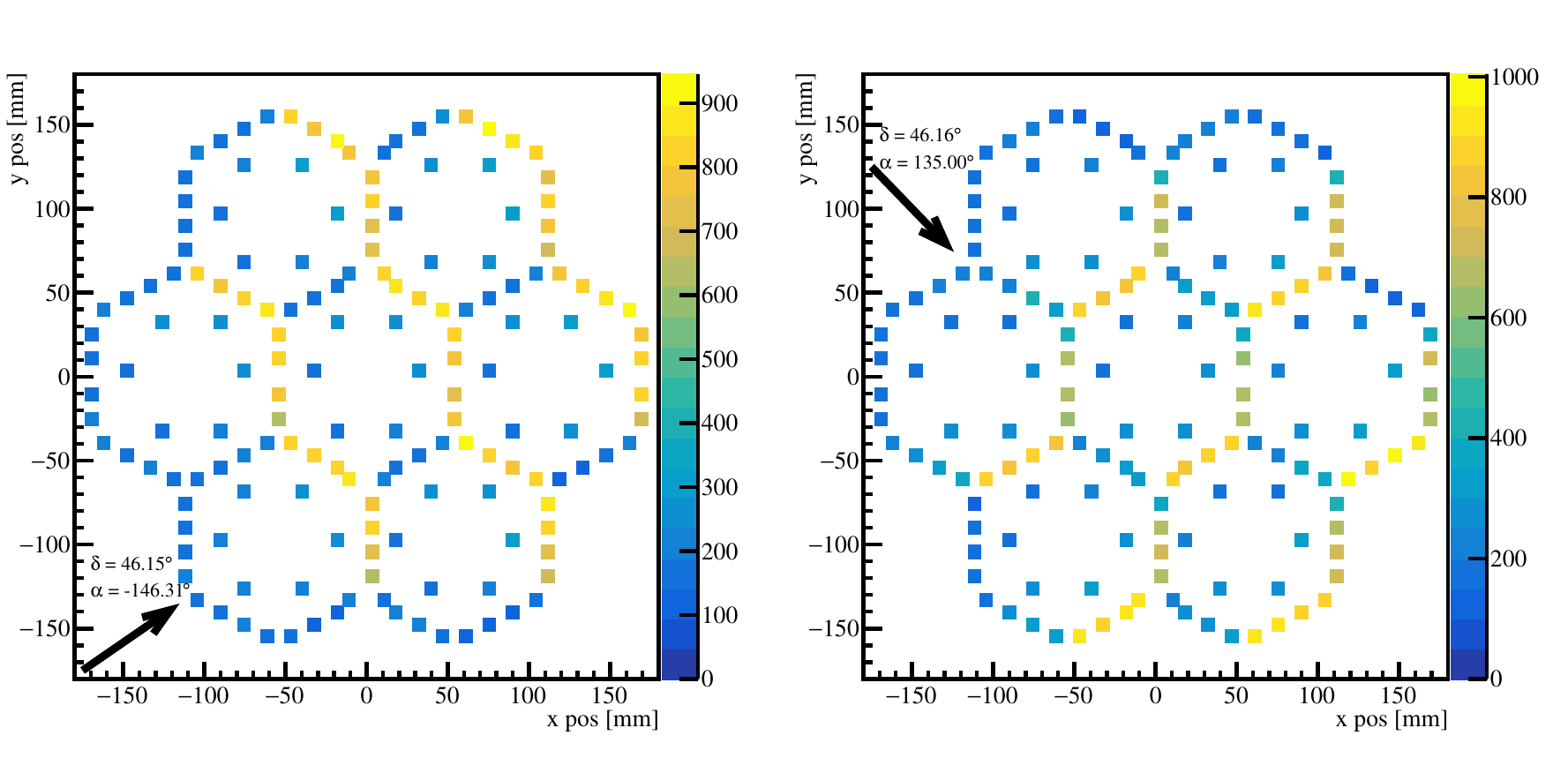}
	\\
	\hspace{1.4 cm}(a)   $\bcol{\delta}=46.1^\circ,\,\bcol{\alpha}=-146.3^\circ$ \hspace{4.5cm} (b)
	$\bcol{\delta}=46.1^\circ,\,\bcol{\alpha}=135^\circ$
	\end{tabular}
	\caption{Variation in counts in different detector units with different
	source positions. Each detector unit is represented by a single point
placed at the center of the unit position.}
	\label{fig:democts}
\end{figure*}

\section{Localisation principle}\label{sec:prin}
Hard X-rays interacting in segmented detectors generate an interaction pattern
dependent on the incidence direction of photons. This can be used to obtain the
photon source direction. 
Previous GRB missions have either used coded masks
\citep{swiftbat,czti} or physically separated detector units oriented in different directions
\citep{fermigbm,batse} to obtain distinct interaction patterns.

SPHiNX records direct interactions (single-hit events) as well as scattered interactions (multi-hit events) from incoming photons. 
Using scattered photons to localise the GRB is difficult as the initial photon direction information is
smeared when the photon scatters. Single-hit events preserve the photon direction information and can be
used to determine the source location.
A GRB observed by SPHiNX will result in a distribution of single-hit counts in
\bcol{the} detector units dependent on the relative location of the GRB. Such a
detected count map is illustrated in
Fig.~\ref{fig:democts}. The
GRB location on sky can then be determined by inverting this count map.
SPHiNX \bcol{is sensitive to the}
photon direction as each detector unit is shielded to different
extents (by the metallic shield and by other detector units).
The use of a flat geometry 
reduces sensitivity to the \bcol{zenith} angle of incident photons (especially for
a source at \bcol{a} large off-axis position).

This paper focuses on exploring three
different inversion routines to obtain the GRB position from the  count
map. 
To simplify the problem, all inversion routines 
assume that a majority of the
detected counts come from a single point source in the sky, with
counts from other sources forming a part of the diffuse sky background. This is a reasonable assumption since the GRB is momentarily
the brightest source in the sky.

The source position is defined using the azimuth \bcol{ ($\alpha$)} and \bcol{zenith
($\delta$)} angles of the source with
respect to SPHiNX. 
Uncertainties in the obtained source position are a result of
assumptions in the inversion routines and Poisson fluctuations in
the detected counts. Each
routine gives a set of estimated position angles \bcol{($\hat{\alpha}$ and $\hat{\delta}$)},
uncertainties \bcol{($\sigma_\alpha$ and $\sigma_\delta$)} and offsets \bcol{ ($\partial_\alpha = |\alpha - \hat{\alpha}|$ 
and $\partial_\delta = | \delta - \hat{\delta} |$)} from the actual position. The offsets and uncertainties are
used later to compare different routines. The polarisation sensitivity is dependent
mainly on \bcol{$\sigma_\delta$}, with
$\bcol{\sigma_\delta} \lesssim$ 5$^\circ$ required to 
minimise the effects of increased MDP in SPHiNX (see Fig.~\ref{fig:vic}).
The hexagonal design results in a high degree of azimuthal symmetry
and reduces dependence of $M_{100}$ on the azimuthal angle. Thus,
polarisation sensitivity is mostly independent of \bcol{$\sigma_\alpha$} and any errors caused
by this weak dependence can be added as systematic errors on polarisation
values as proposed in \citet{gapbrst}.

\section{Routine implementation and results}\label{sec:uncer}
The routines perform an inversion by first obtaining a mapping function between
detected counts and sky position. This function is approximated by an empirical
relation in
one routine and by a pre-computed lookup table in the other two routines.

\begin{figure}[!ht]
	\centering
	\includegraphics[width=0.5\textwidth]{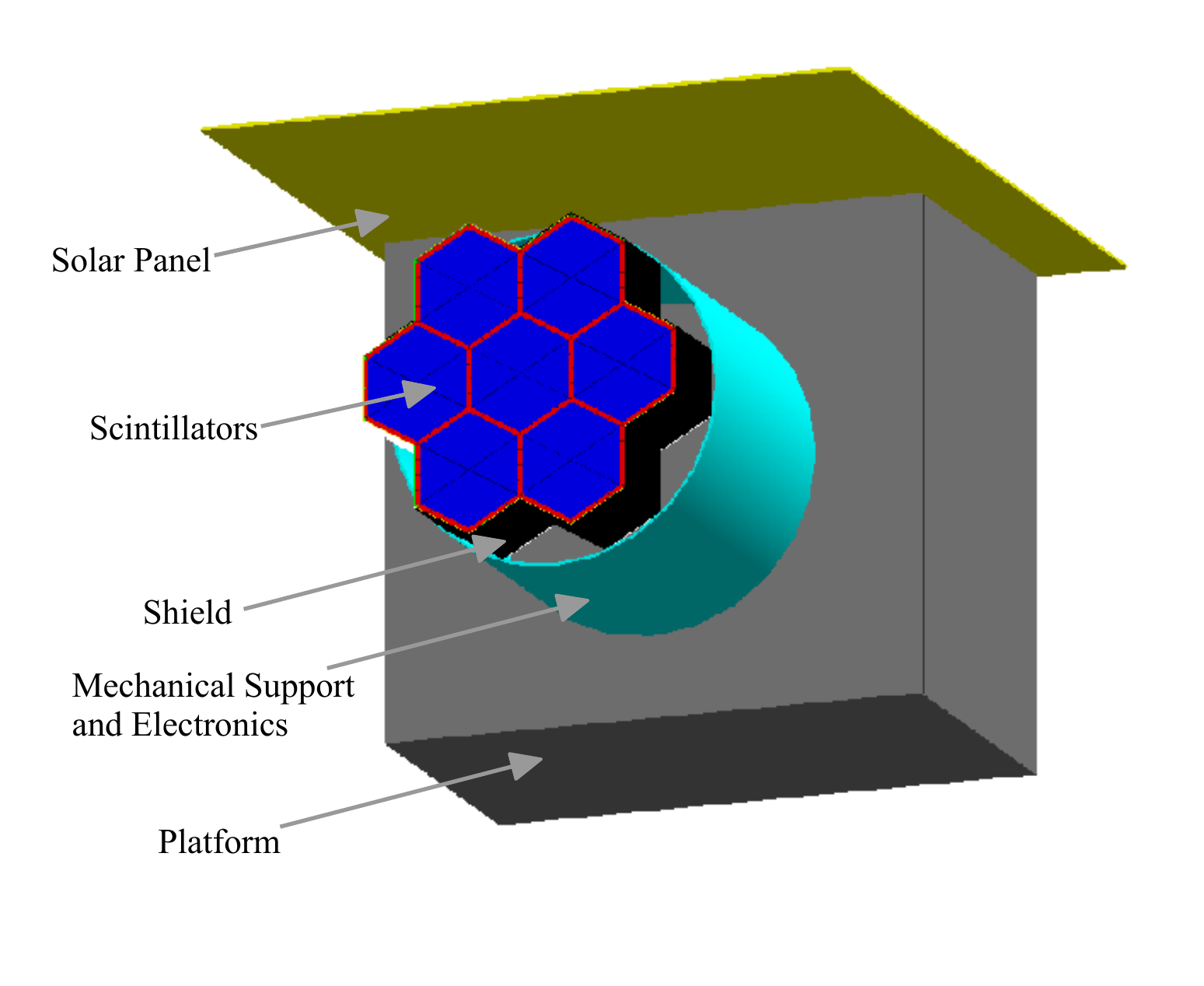}
	\caption{View of the model used for Geant4 simulations. All components such
	as the satellite platform, shielding materials, scintillators, electronic
	read-out boards and solar panel have been included in the model.}
	\label{fig:g4model}
\end{figure}

In order to evaluate the routines (and compute the lookup table), a Monte-Carlo simulation of the SPHiNX
polarimeter based on Geant4 (version 10.02.p02) \citep{geant4} is used. Details of the
simulation set-up (physics lists, mass model and event selection logic) are
given in Refs~\citenum{fxie18,pearce18}. The Geant4 model used for this simulation is
shown in Fig.~\ref{fig:g4model}. \bcol{Simulated photons are emitted from a
disk of radius 20 cm located at different positions in the SPHiNX field of
view.}
ROOT \citep{root2} is used both for handling the
output data from Geant4 and for implementing the routines to perform the inversion.
The data stored from each simulation run consists of \bcol{hit information of
each event such as deposit energy, incident photon direction, detector unit in
which the hit occurred and incident energy} stored in ROOT Trees.  

The inversion routines are evaluated by simulating a GRB in the SPHiNX
field of view. GRB-120107A (with Band parameter $\alpha=-0.94$, $\beta=-2.39$ 
and $\mathrm{E_p}=201.22$ keV) is chosen as a representative
GRB for testing the inversion routines. The choice of this representative GRB is 
not critical as effects of differing GRB spectra are corrected for (see
Section~\ref{ss:reschi}). Effects of changing fluence and position (in sky)  are
evaluated by simulating the
representative GRB with the parameters shown in Table~\ref{tab:pos}. 

\begin{table*}[!h]
\centering
\caption{Fluence and position \bcol{($\alpha,\delta$)} values for evaluating routines}
\label{tab:pos}
\begin{tabular}{|c|c|c|p{5cm}|}
\hline
\textbf{Fluence} & \textbf{\bcol{$\alpha$}} & \textbf{\bcol{$\delta$}} & \textbf{Remarks} \\
\hline
2 ph/cm$^2$ to 200 ph/cm$^2$ & -54$^\circ$ & 32.8$^\circ$ & Evaluates the
effect of GRB fluence (in 50-300 keV) for weak (2~ph/cm$^2$), median (20~ph/cm$^2$)
and very strong GRBs (200~ph/cm$^2$). \\ \hline
14.5 ph/cm$^2$ & -180$^\circ$ to 180$^\circ$. & $43.7^\circ$ & Evaluates the effect of azimuthal angle. \\ \hline
14.5 ph/cm$^2$ & 33$^\circ$ &  0$^\circ$ to 90$^\circ$ & Evaluates the effect of
\bcol{zenith} angle. \\
\hline
\end{tabular}
\end{table*}

\subsection{Database simulations}\label{ss:methods}
The lookup table mapping the detected counts to sky position is maintained
in the form of a database. The database
consists of a set of count maps generated via simulations, for a large number of
sky positions. This is a commonly used tool for numerically approximating the mapping
function. It minimises computations by having the mappings pre-computed and
stored. \textit{Fermi} GBM uses a database with 41,168 points on an equi-spaced two-dimensional sky grid obtained by
linearly interpolating between 272 simulated points \citep{locgbm2} to obtain
their database. \textit{POLAR}~\citep{Polar} uses a database with 10,201 simulated points on a
two-dimensional planar Cartesian grid with no interpolations as their database
\citep{locpol}.

\begin{figure}[!ht]
	\centering
	\includegraphics[width=0.49\textwidth]{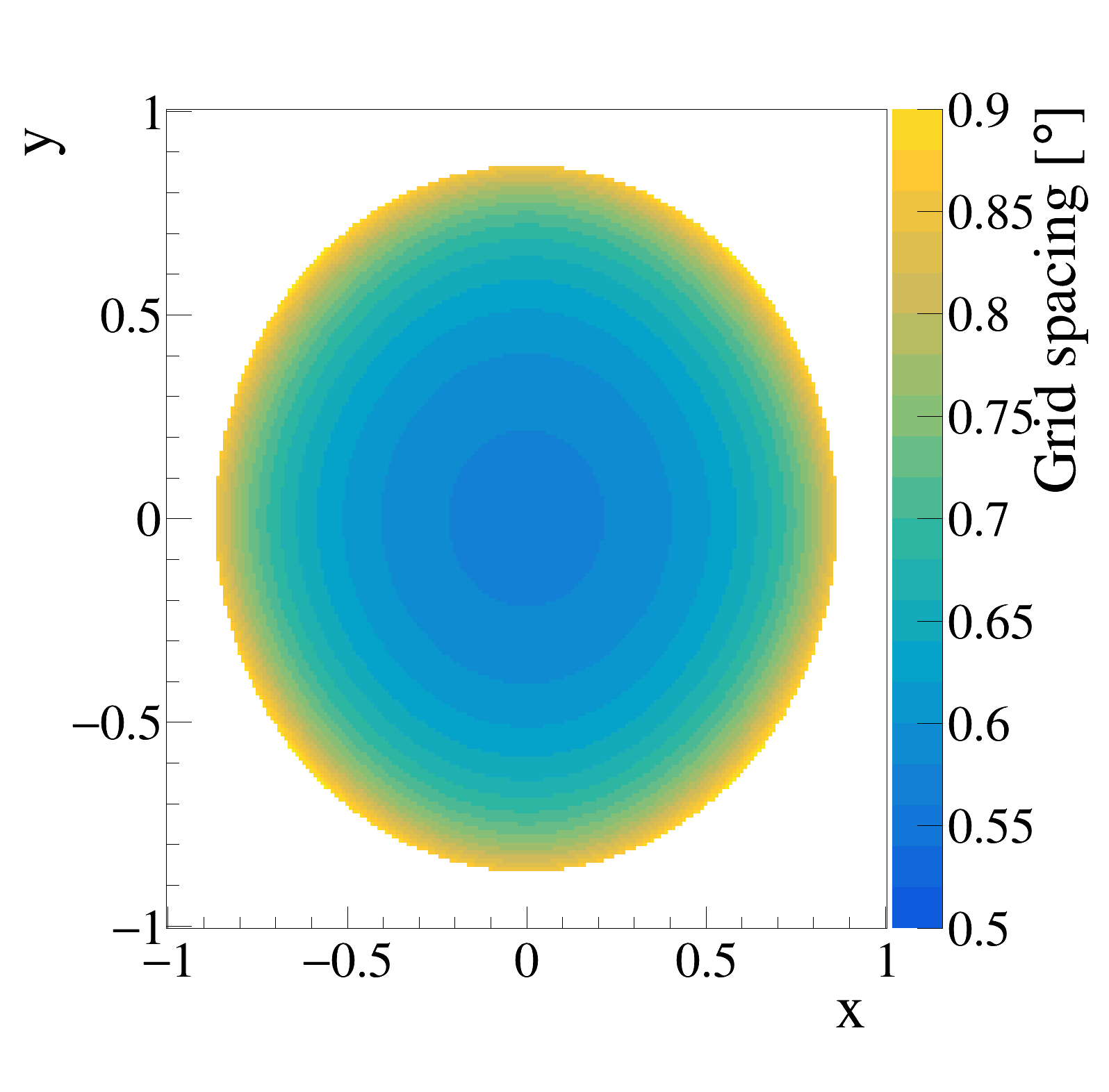}
	\caption{ Grid spacing (in degrees) as a function of sky position. The
		outer circle in figure corresponds to the edge of FOV (120$^\circ$) and
the centre marks position of on-axis photons.  }
	\label{fig:grid}
\end{figure}

For SPHiNX, 31,730 points are chosen on a regular Cartesian grid similar to
\textit{POLAR}. The transformation between this grid ($x$,$y$) and the sky
co-ordinates \bcol{($\alpha$,$\delta$)} uses the basic mapping between spherical and Cartesian
co-ordinates. The grid \bcol{runs from -1 to 1} along each of $x$ and $y$
axes \bcol{with a spacing of 0.01 between grid points in both directions.
This results in an} angular separation between grid positions as seen in
Fig.~\ref{fig:grid}. \bcol{A Cartesian grid is used since uncertaintes 
can be conveniently determined using interpolation on the regularly spaced
rectangular grid (see
Section~\ref{ss:reschi}).}

To make the database, steps given in Algorithm~\ref{alg:dbase} are used. 
The use of a very bright GRB (200~ph/cm$^2$) minimises statistical fluctuations in the database 
(and thereby in the mapping function). The spectrum is flat
(equal flux) across all energies to enable spectral corrections as described in
Section~\ref{ss:reschi}. The effect of these fixed
values (of spectrum and fluence) used to generate the reference database is discussed further in
Section~\ref{ss:ressys}. \bcol{To scale for the source flux, the database stores
fractional counts as shown in Eqn.~\eqref{eqn:mni}, where the terms are
as defined in Algorithm~\ref{alg:dbase}.}
\begin{equation}
	\bcol{\bar{m}_i(x,y)} = \frac{m_i(x,y)}{\sum\limits_i m_i(x,y)}
	\label{eqn:mni}
\end{equation}

\begin{algorithm}[h!]
\renewcommand{\thealgorithm}{A}
\caption{Database generation}
\label{alg:dbase}

\begin{enumerate}
	\item Select a very bright GRB to generate \bcol{the} database. Fluence of 200
ph/cm$^2$. Flat spectrum. 
\item Step $x$ from \bcol{$-1$ to $1$ at a step size of $0.01$}. 
\item Step $y$ from \bcol{$-1$ to $1$ at a step size of $0.01$}.
\item If \bcol{$x^2 + y^2 > 1$}, then reject that point.
\item Run Geant4 simulation for each combination $x,y$. 
\item Store all event and hit information. This forms 
	the extended database.
\item Store direction information along with original and deposited energies,
	of single-hit events only (with deposited energy $> 50$ keV and $<$ 600 keV). This forms the
	reduced database.
\item Get the detected count map $m_i(x,y)$ (for all detector units $i$) and
	obtain \bcol{the fractional counts}
	$\bcol{\bar{m}_i(x,y)}$ as given in Eqn.\eqref{eqn:mni}. Store $\bcol{\bar{m}_i(x,y)}$ for all
	positions ($x,y$) to form the truncated database.
\end{enumerate}
\end{algorithm}

The entire extended database (including details of all hits and
events) takes $\sim$ 1.2 Terabytes of disk space, while the reduced database (with
details of single-hit events and necessary ancillary information) takes $\sim$
80 Gigabytes space. \bcol{The thresholds used for single-hit event selection (see
Algorithm~\ref{alg:dbase}) in the reduced database are chosen to reduce counts from the diffuse
background.} While a truncated database with just $\bcol{\bar{m}_i(x,y)}$ values
(500 Megabytes) can also be used, this does not allow for GRB spectral
corrections.  

\subsection{Routine-I : Modulation curve}\label{ss:resmod}
The modulation curve method  maps the counts
in the outermost detector units to source position using an empirical relation
composed of two sinusoidal components with 180$^\circ$ and 360$^\circ$
periodicities (see Eqn.~\eqref{eqn:modln}). \bcol{The outermost detector units
(consisting of GAGG units on the periphery of the entire detector array) are used as their position 
makes the detected counts sensitive to source
position. The 180$^\circ$ component arises because the GAGG units are
shaped like rectangular slabs, where some units have their long edge facing the
GRB direction and other units (placed 90$^\circ$ away) have the short edge
facing the source direction. Thus the detected counts are modulated in
proportion to the geometrical cross section of the GAGG units as seen by the
GRB. The 360$^\circ$ component arises because only one side of the 
GAGG walls (the outer side) is covered by a
multi-layered shield which reduces counts for a source placed on that side. 
Thus, GAGG units placed on the opposite side of the detector array as seen from
the GRB will have higher counts than GAGG units placed closer
to the source. The sum of these two components describes the count modulation in
the periphery units.}

The phase ($p_1$ and $p_4$) of the modulation in Eqn.~\eqref{eqn:modln}  can
be used to determine the azimuthal angle \bcol{($\alpha$)}, while the amplitude ($p_0$
and $p_3$) can be used to determine the \bcol{zenith} angle \bcol{($\delta$)}. The mapping
makes use of detector angle $\phi_d$, defined as the angle subtended by the centre 
of each detector unit, at the centre of the local hexagonal (see Fig.~\ref{fig:scat}(a)).
The steps involved in implementing this routine are specified in
Algorithm~\ref{alg:modln}.  

\begin{equation}
f(\phi_d) = p_2 + p_0 \cdot cos(2(\phi_d+p_1)) + p_3 \cdot cos(\phi_d+p_4),
\label{eqn:modln}
\end{equation}

\begin{algorithm}
\renewcommand\thealgorithm{I}
\caption{Modulation curve}
\label{alg:modln}	

\begin{enumerate}
\item Compute $\phi_d$ for all detectors on the periphery. This gives 24
	possible angles (bins) for 72 detector units (such that 3 detector units have the same
	angle $\phi_d$). 
\item Average the counts of the 3 units per angular bin. Plot averaged counts
	against the angle $\phi_d$.
\item Fit Eqn.\eqref{eqn:modln} to the detected counts and obtain parameters $p_0$ to
	$p_4$. To weight the fit, use the propagated error obtained from Poisson
	variances in counts of each detector unit (see Fig.~\ref{fig:mcurve}).
\item Use relation in Eqn.\eqref{eqn:mc_hphi} to estimate $\bcol{\hat{\alpha}}$.
\item Find $a_m$ using \bcol{$\hat{\alpha}$}. Use Eqn.\eqref{eqn:mc_hthet} 
	to estimate \bcol{$\hat{\delta}$}. 
\item Propagate errors on the fit parameters to obtain the uncertainty
	\bcol{($\sigma_\alpha , \sigma_\delta$)} on estimated values.
\end{enumerate}
\end{algorithm}

\begin{align}
	\bcol{\hat{\alpha}} &= \label{eqn:mc_hphi}
	\begin{cases}
		p_4 - \pi \, ; \mathrm{Case\,I} \\
		p_1  \,; \mathrm{Case\,II}
	\end{cases} 
	\\
	\bcol{\hat{\delta}} &= \cos^{-1}\left(\frac{p_0}{a_m}\right) \label{eqn:mc_hthet} 
\end{align}

\begin{figure}[!ht]
	\centering
	\includegraphics[width=0.5\textwidth]{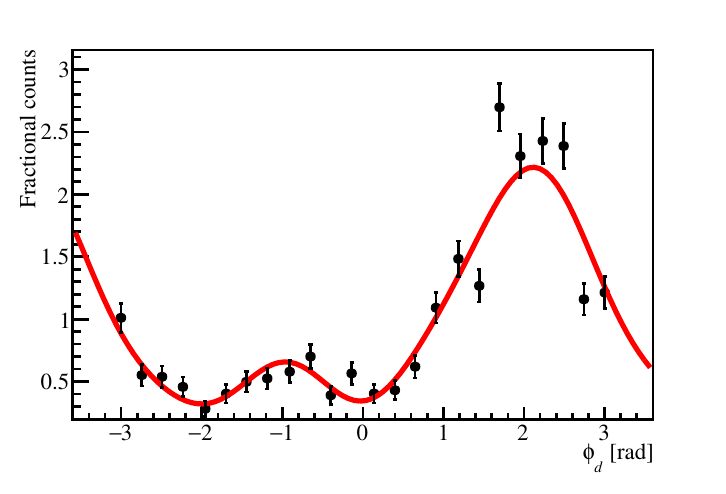}
	\caption{Example modulation curve of outermost detectors (for
		\bcol{$\alpha=-63^\circ , \delta=48^\circ$)} fit by
	Eqn.~\eqref{eqn:modln}}
	\label{fig:mcurve}
\end{figure}

The \bcol{relationship} between parameters ($p_0$ to $p_4$) and sky position
\bcol{($\delta$,$\alpha$)} is
obtained by searching for correlations in randomly simulated GRBs at
different sky positions. It is seen that the phase
of both 360$^\circ$ ($p_4$) and 180$^\circ$ ($p_1$) components correlate well with the
azimuthal angle \bcol{($\alpha$)}. This correlation is independent of
\bcol{zenith angle ($\delta$)} and
the source spectrum.  The 360$^\circ$ component maps increased counts with
detector units opposite to the azimuthal direction of the source. The
180$^\circ$ component maps increased counts on both the detector units facing the GRB
as well as on detector units placed diametrically opposite. Thus the
phase angles $p_1$ and $p_4$ have similar values and either of these
can be used to find \bcol{$\alpha$}. However, for some source positions
(at low \bcol{$\delta$} values) the 180$^\circ$ component maps these modulations better
(as the amplitude of the 360$^\circ$ is smaller than the 180$^\circ$ for small
\bcol{$\delta$}). 
Thus, the preferred way to obtain \bcol{$\hat{\alpha}$} is to use $p_4$ directly 
when $p_1$ and $p_4$ match (Case I). 
When they do not match (Case II), the phase angle indicated by $p_1$
is used. 

\begin{figure}[!ht]
	\centering
	\includegraphics[width=0.5\textwidth]{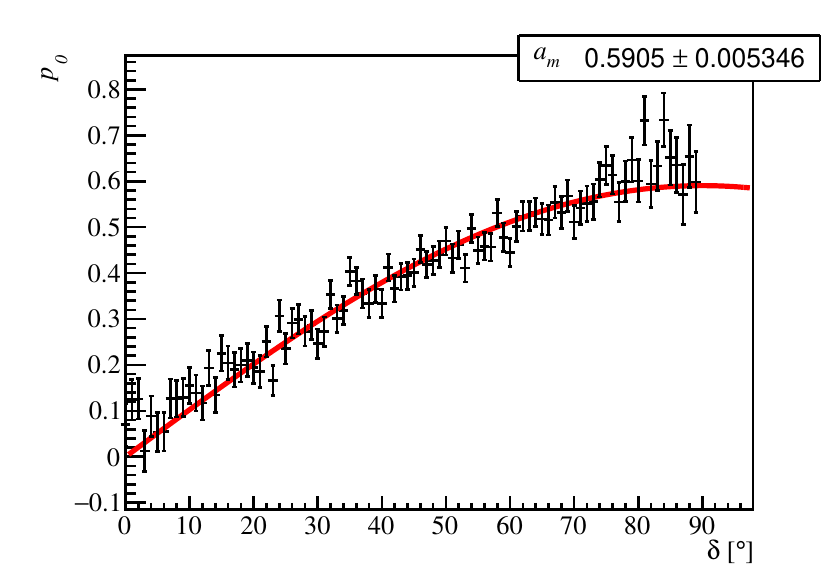}
	\caption{Relation between angle \bcol{($\delta$)} and amplitude $p_0$ (see
	Eqn.~\eqref{eqn:mc_hthet})}
	\label{fig:mthet}
\end{figure}

As mentioned before, the amplitude of the 360$^\circ$ component is sensitive to
the \bcol{zenith} angle. \bcol{This is because the modulation
will be minimum (ideally zero) when the source is at the zenith (all units
are equally illuminated) and maximum when the source is near the horizon (units
facing the source are maximally illuminated).}
Fig.~\ref{fig:mthet} shows \bcol{this relation between \bcol{$\delta$} and amplitude $p_0$
	as obtained from the simulations. The sinusoidal relation, with parameter
$a_m$, is quantified by Eqn.~\eqref{eqn:mc_hthet}}. The expression has a weak
dependence on \bcol{$\alpha$} (since $a_m$ changes slightly with \bcol{$\alpha$)}. Thus, a two-step
solution is needed to find the source location. Values of $a_m$ for different
\bcol{$\alpha$} are pre-computed using simulations and are seen to vary by less than 10\%. 

\begin{figure*}[!ht]
 \centering
 \begin{tabular}{l}
	\includegraphics[width=0.97\textwidth]{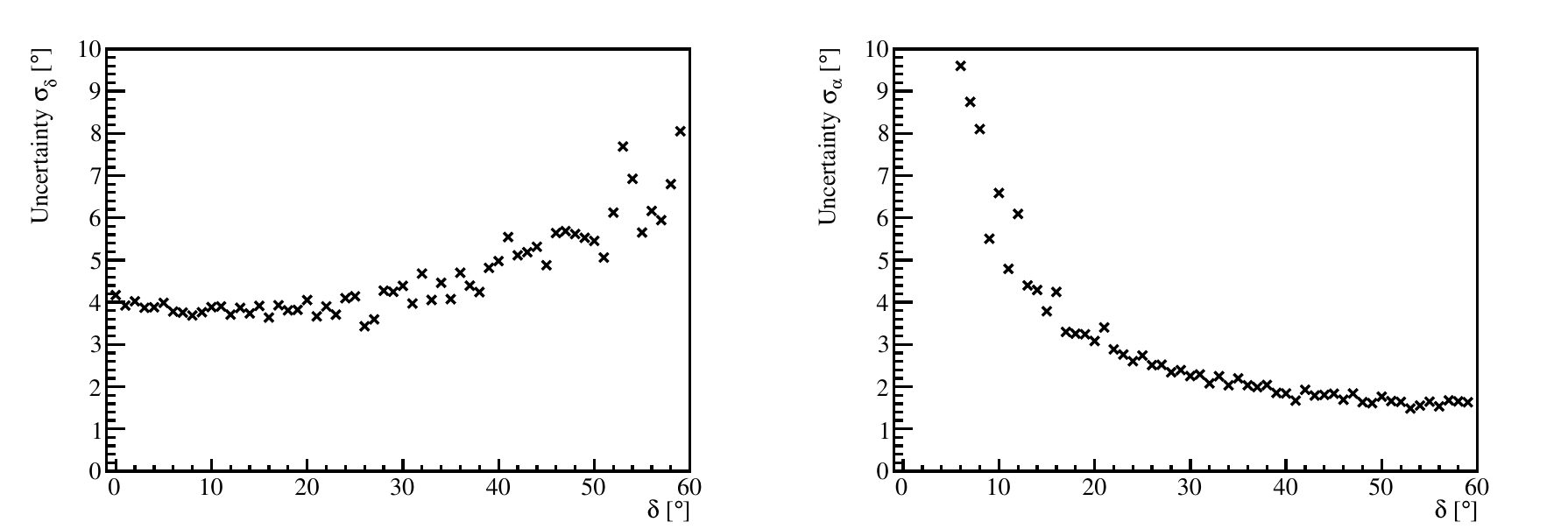}
	\\
	\hspace{4.4cm}(a)    \hspace{8.5cm} (b)
	\end{tabular}
	\caption{Uncertainties (a) \bcol{$\sigma_\delta$  and (b) $\sigma_\alpha$} as a function of
		\bcol{zenith angle ($\delta$)} for the modulation curve routine. The values are
	plotted till $\bcol{\delta}=60^\circ$ (edge of FOV) for a GRB of median fluence.}
	\label{fig:resmod}
\end{figure*}

\bcol{$\sigma_\delta$ and $\sigma_\alpha$} are evaluated for different parameters as specified in Table~\ref{tab:pos} (see Fig.~\ref{fig:resmod}). For
large \bcol{$\delta$, $\sigma_\delta$}
increases due to low counts in many detector units. At
small \bcol{$\delta$}, the position angle \bcol{$\alpha$} is not very well defined,
thereby leading to larger \bcol{$\sigma_\alpha$}. As seen in following sections, these trends are true irrespective of the
localisation routine used. The important result worth noting is that
$\bcol{\sigma_\delta} \lesssim 8^\circ$ at all positions within the FOV (for a median fluence GRB).

\subsection{Routine-II - Minimisation of $\chi^2$}\label{ss:reschi}
The minimisation of Pearson's $\chi^2$ is a classic routine 
used by \textit{Fermi}
GBM\citep{locgbm2}, \textit{CGRO} BATSE \citep{locgbm},
\textit{POLAR}~\citep{locpol}, \textit{AstroSat} CZTI~\citep{czti}, SSM~\citep{ssm} and many other 
indirect imaging missions. 
The mapping function is obtained numerically and makes use of a pre-computed
database. 
The inversion is done by \bcol{comparing the measured counts ($c$) with the
database modeled counts ($m$) using the} $\chi^2$ statistic
(Eqn.~\eqref{eqn:chimod}). \bcol{This statistic is computed} for each position in the simulated database 
and \bcol{the location of the minimum of the statistic is chosen as the best
estimate of the source position.}
\begin{equation}
	\chi^2(x,y) = \sum_{i=1} \frac{(c_i-c_{tot}\cdot
	\bcol{\bar{m}_i(x,y)})^2}{c_{tot}\cdot \bcol{\bar{m}_i(x,y)}} ,
	\label{eqn:chimod}
\end{equation}
where, $c_i$ is the measured counts in detector unit $i$, $c_{tot} = \sum c_i$ is total number of counts in the
count map, and $\bcol{\bar{m}_i(x,y)}$ the \bcol{fractional} database counts.
The estimation of uncertainties is based on the $\chi^2$ density distribution
for binned count data, which in turn uses the assumption of Gaussian distributed
data in each detector unit (bin). The steps used to implement this routine are given in
Algorithm~\ref{alg:chi2}. 

\begin{algorithm}[h!]
\renewcommand\thealgorithm{II}
\caption{$\chi^2$ routine}	
\label{alg:chi2}
\begin{enumerate}
	\item Obtain spectrum  of the GRB. Get corrected database count for this
		spectrum using Eqn.\eqref{eqn:corrspec}. 
	\item Compare detected count map with database using
		Eqn.\eqref{eqn:chimod}. 
	\item Chose position with a minimum $\chi^2(x,y)$ value ($\chi^2_{min}$) as 
		source position $(x_{min}, y_{min} )$. 
	\item Estimate \bcol{$\hat{\delta}$ and $\hat{\alpha}$} using position of minimum and
		co-ordinate transformation ($x_{min},y_{min}~\mapsto~\bcol{\hat{\delta},\hat{\alpha}}$).
	\item Get variation of $\chi^2$ with $x$ (at $y=y_{min}$) around minimum.
		Fit this with a parabola. Find $x_1$ and $x_2$ at which the parabola
		intersects $\chi^2 = \chi^2_{min} + 1$. Get $\sigma_x = 0.5(x_1 - x_2)$
	\item Repeat procedure to get $\sigma_y$.
	\item Find \bcol{$\sigma_{\delta}$ and $\sigma_{\alpha}$} using $\sigma_x,\sigma_y$ and
		error propagation (from Cartesian to spherical co-ordinates).
\end{enumerate}
\end{algorithm}

The counts $m_i$ are corrected for spectral shape using Eqn.\eqref{eqn:corrspec}. 
These corrected $m^\prime_i$ are then used to get the \bcol{fractional} counts
$\bcol{\bar{m}_i}$. Correction for the source spectrum is done by dividing the SPHiNX energy range
into 5 keV bands. In Eqn.\eqref{eqn:corrspec}, $m_{ij}$ is the detected counts in the
$j$-th (5 keV) energy bin for detector unit $i$ and
$k_j$ is the spectral flux in the $j$-th band. $N^{uni}$ is the number of integrated
photons across the flat spectrum  (used to generate the database) and $N^{GRB}$ denotes the integrated photons
across the actual GRB spectrum. In order to perform this correction, the database
needs to store energy information of each simulated hit. This makes the database
\bcol{relatively} large in size. 
The correction in Eqn.\eqref{eqn:corrspec}
assumes that counts in a particular band are only affected by photons in that
band. \bcol{This ignores the detector response which will lead to an
additional systematic uncertainty as discussed in
Section~\ref{sec:discussion}.} 

\begin{equation}
	m^\prime_{i} = \frac{N^{uni}}{N^{GRB}} \sum_j \left(m_{ij} \cdot \frac{ 
	k_j^{GRB} } {k_j^{uni} }\right).
\label{eqn:corrspec}
\end{equation}

\begin{figure*}[!ht]
	\centering
	\begin{tabular}{l}
	\includegraphics[width=0.97\textwidth]{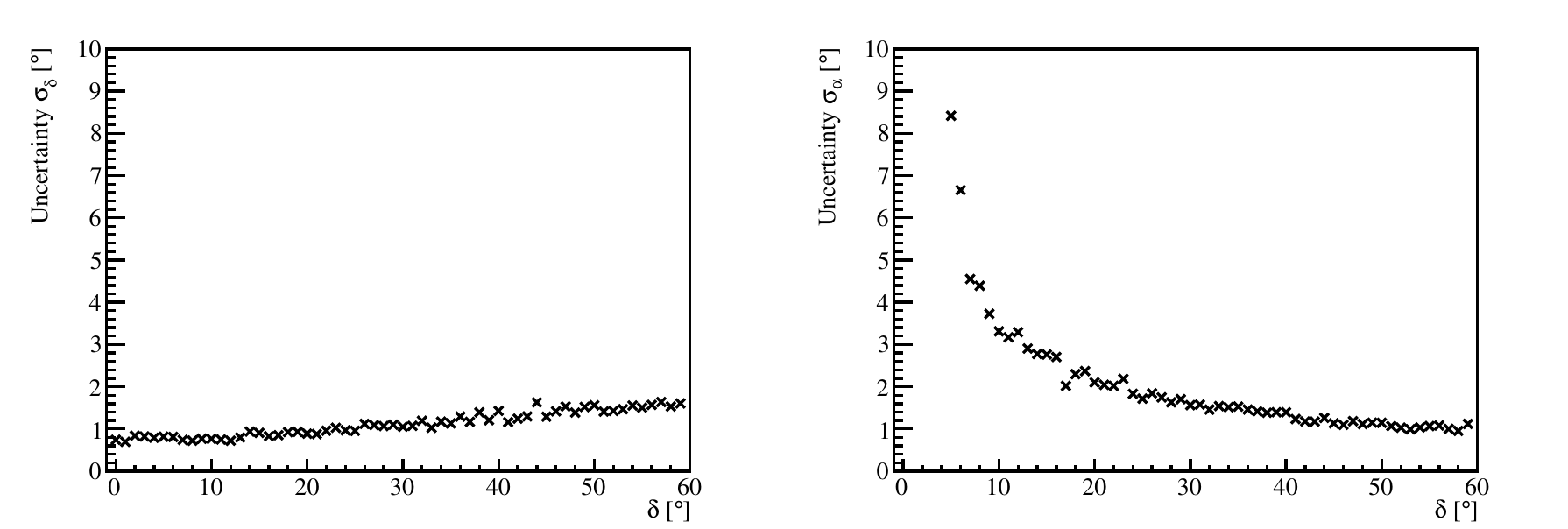}
	\\
	\hspace{4.4cm}(a)    \hspace{8.5cm} (b)
	\end{tabular}

	\caption{Uncertainty (a) \bcol{$\sigma_{\delta}$} and (b)
		\bcol{$\sigma_\alpha$} for $\chi^2$
	routine as a function of \bcol{$\delta$}.}
	\label{fig:reschi}
\end{figure*}

Fig.~\ref{fig:reschi} shows the obtained uncertainties as a
function of source position \bcol{$\delta$} for this routine (for a median fluence GRB). As
seen, the obtained uncertainties are much lower than those for the modulation curve
routine. The trends though, remain the same as Routine-I :
\bcol{$\sigma_\delta$} increases with \bcol{$\delta$} and \bcol{$\sigma_\alpha$} has very high values for low \bcol{$\delta$}. 

Instead of looking at uncertainties on two separate parameters, it is convenient to express
localisation uncertainty in terms of the error radius. 
The error radius is \bcol{defined in terms of the half angle of a cone which
subtends the same solid angle on the sky as the uncertainty region defined by
\bcol{$\sigma_\delta$} and \bcol{$\sigma_\alpha$}. This means that the error
radius is well defined even when the azimuthal angle (\bcol{$\alpha$}) is
undefined near zenith. For small uncertainties,} \bcol{$\sigma_\delta$}
and  \bcol{$\sigma_\alpha$} can be converted to the error radius
\bcol{($\psi$)} using
\begin{equation}
	2\pi\cdot\big(1 - \bcol{\cos(\psi)}\big) \approx \bcol{4 \cdot \sigma_\delta} \cdot
	\bcol{\sigma_\alpha} \cdot \sin{\bcol{(\hat{\delta})}}
	\label{eqn:errad}
\end{equation}

\begin{figure}[!h]
	\centering
	\includegraphics{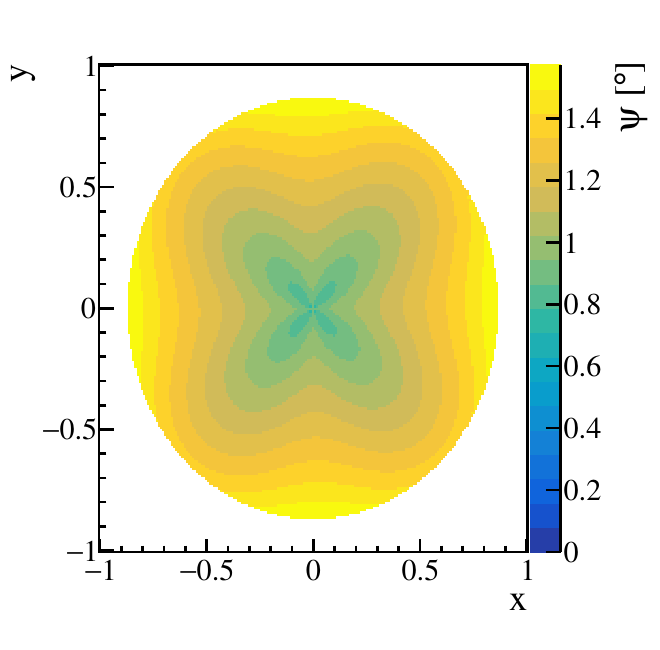} 
	\caption{Error radius as a function of sky position. Figure
		obtained using Eqn.~\eqref{eqn:errad} with
		\bcol{$\sigma_\alpha,\sigma_\delta$} obtained at
	positions listed in Table~\ref{tab:pos}.}
	\label{fig:errad}
\end{figure}

Fig.~\ref{fig:errad} shows the error radius for the $\chi^2$ routine at different positions on the sky. The error radius
increases towards the edge of the FOV and is less than $\sim 1^\circ$ for most
of the FOV. While the obtained uncertainties are very low, these values do not account for systematics and background
effects, which are discussed further in Section~\ref{sec:discussion}.

The error radius can also be directly obtained using two-dimensional contours on the
$\chi^2$ value. \bcol{However, the
grid-spacing used is often too sparse to get smooth
contours as discussed in Section~\ref{sec:discussion}. The simpler procedure of
using parabolic fits is therefore preferred for the $\chi^2$ routine.}

\subsection{Routine-III : Maximising the likelihood\label{ss:reslik}}
The implementation and details of the likelihood routine are very similar to the
$\chi^2$ routine. The differences arise in the method used to get the best
estimate, in the calculation of uncertainties and
in the use of a prior. The steps used to implement the likelihood routine are
given in Algorithm~\ref{alg:rout3}.
The $\chi^2$ routine provides a convenient approach for
performing the inversion in the limit of Gaussian distributed data, which 
holds true when sufficiently high ($\gtrsim 20$) counts are present in all
detector units. This is indeed true for many GRBs \bcol{(with fluence above median
GRBs)}. For the other GRBs a more accurate routine involves computing the
Poisson based likelihood function and maximising this likelihood for obtaining
the inversion.

\begin{algorithm}[h!]
\renewcommand\thealgorithm{III}
\caption{Likelihood Routine}
\label{alg:rout3}
\begin{enumerate}
	\item Correct database counts for the GRB spectrum
		(Eqn.~\eqref{eqn:corrspec}).
\item Compute 2D posterior probability in $(x,y)$
	with a uniform prior for GRBs in the sky using Eqn.~\eqref{eqn:bayes2}.
\begin{enumerate}
	\item For each point $(x,y)$ in sky compute likelihood that detector units
		$i$ will observe the counts $m_i(x,y)$ and find its logarithm. Add 
		the log-likelihoods for all detector units $i$ to find total
		log-likelihood.
	\item Take the exponential of the total log-likelihood and multiply it with the prior
		$\left(\sqrt{\frac{x^2 + y^2}{R^2}}\right)$
\end{enumerate}
\item Make a regular grid in \bcol{$(\delta,\alpha)$}. Linearly interpolate on the database grid to find 
	posteriors for all the points on the new grid.
\item Obtain one-dimensional posterior probability distributions over \bcol{$\delta$}
	and \bcol{$\alpha$} separately by marginalising this two-dimensional distribution. 	
\item Obtain an estimate of the source position \bcol{($\hat{\delta},\hat{\alpha}$)} using the MAP (maximum
	a posteriori estimate) with uncertainties \bcol{($\sigma_\delta,\sigma_\alpha$)} computed using the 68\% confidence
	interval around the MAP estimate.
\end{enumerate}
\end{algorithm}

The Poisson based likelihood function, \bcol{which relates the measured counts
$m$ to the database modeled counts $c$,} is easily computed using the expression
for the Poisson distribution as 
\begin{equation}
	L(c_i | \bcol{\delta,\alpha})  = \prod_i \frac{m_i(\bcol{\delta,\alpha})^{c_i}}{c_i !}e^{\bcol{-m_i(\delta,\alpha)}}
	\label{eqn:likel}
\end{equation}

Using Bayesian inference and a uniform prior for distribution of GRBs in the sky gives the posterior
probability distribution $p(x,y | c_i)$ as 

\begin{equation}
\begin{split}
	p(x,y | c_i) \propto \prod_i \left\{ \frac{\big(c_{tot} \cdot \bcol{\bar{m}_i(x,y)}
	\big)^{c_i}}{c_i !} \exp\big(- c_{tot} \cdot \bcol{\bar{m}_i(x,y)}\big)\right\} \\
	\cdot \left(\sqrt{\frac{x^2 + y^2}{R^2}}\right)
	\label{eqn:bayes2}
\end{split}
\end{equation}

\begin{figure*}[!ht]
	\centering
	\begin{tabular}{l}
		\includegraphics[width=0.7\textwidth]{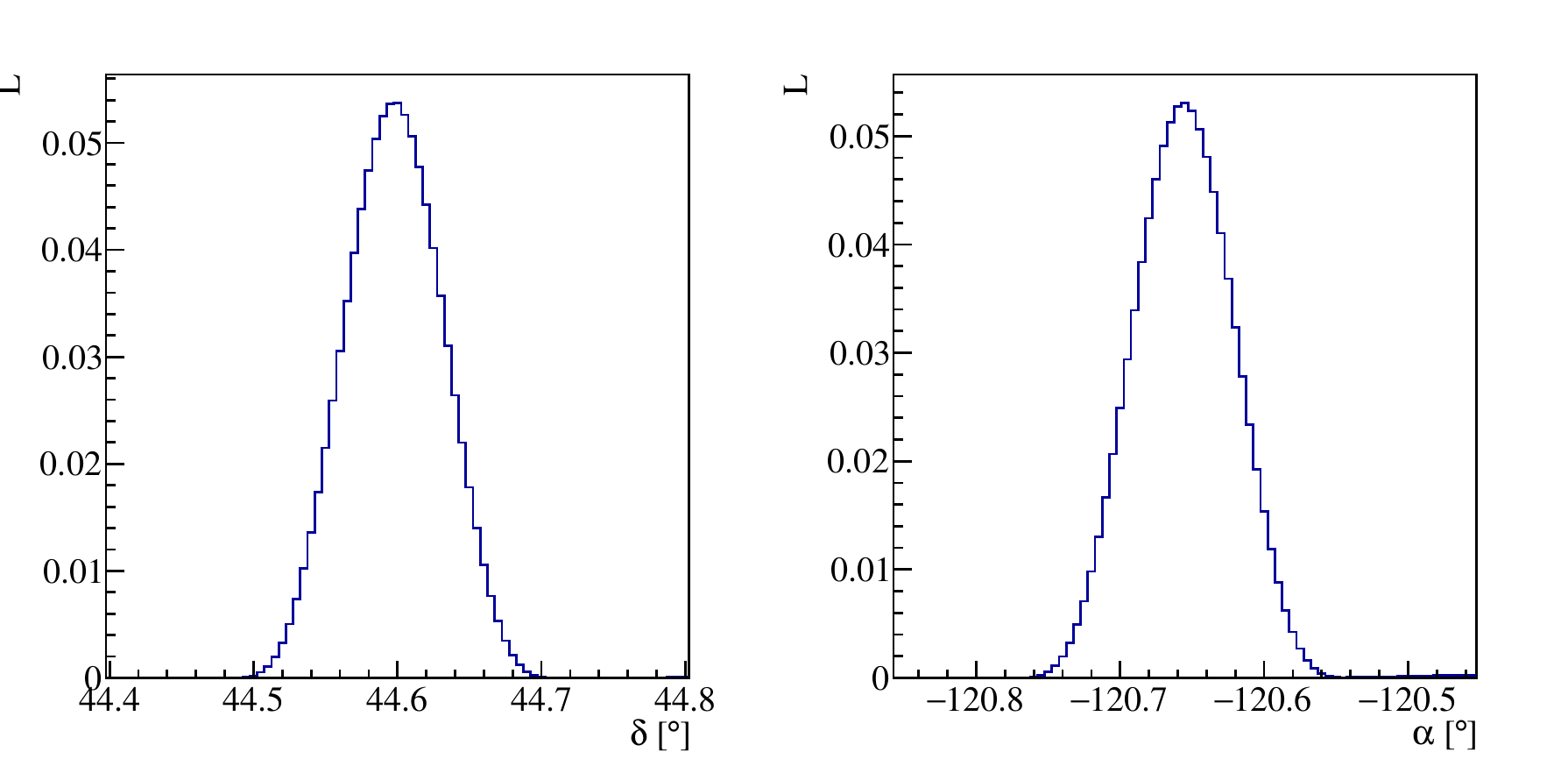}
	\\
	\hspace{2.8cm}(a)   \hspace{4.5cm} (b) 
	\end{tabular}

	\caption{Marginalised posterior in (a) \bcol{$\delta$} and (b) \bcol{$\alpha$} for the likelihood
	routine (for a median fluence GRB)}
	\label{fig:reslik}
\end{figure*}

In this routine, the likelihood is computed for all points in the
database. However, to estimate uncertainties, more points are needed as the
marginalised posterior will often be confined to one or two grid-intervals of the
database. A linear 2D interpolation of the posterior probability is 
done over a regular grid in \bcol{$\delta,\alpha$} for ease of obtaining marginalised posteriors.
A typical posterior distribution for a median fluence GRB obtained
using a finely gridded database  (grid-size \bcol{0.001} in $x,y$) and linear interpolation
(to a 0.05$^\circ$ regular grid in \bcol{$\delta,\alpha$)} is shown in Fig.~\ref{fig:reslik}. 

The figure shows that uncertainties in the likelihood routine can be as low as
$\sim 0.2^\circ$ if a finely gridded database is used. Typically systematic errors
(Section~\ref{ss:ressys}) will be
much larger than this and will dominate the localisation uncertainty .

\section{Summary and Discussions} \label{sec:discussion}
The results obtained demonstrate the ability of the SPHiNX design to
localise GRBs. The effect of systematics like the presence of background
and the use of a gridded database generated via Monte-Carlo simulations is
considered below.  

\subsection{Systematics and Background effects}\label{ss:ressys}
There are a number of additional sources of uncertainty in localisation, which
arise from the design and
characteristics of the detectors used and from approximations
used to implement each routine. The most important 
ones which dominate the uncertainty on localisation are listed below.


\subsubsection{Detector effects (efficiency and energy resolution) }
The routines assume that all detector units have equal efficiency for photon
detection. Calibration of the relative efficiencies of all detector units would be required to
renormalise the obtained counts. Any uncertainty in the detector efficiency calibration will
propagate to uncertainty in source position. 

The two database routines use spectral corrections as given in Eqn.~\eqref{eqn:corrspec}. The
spectral correction depends on the knowledge of the source spectrum, which in
turn depends on the spectral response. GAGG units with a
better energy resolution (than plastic) of $\sim$24\% at 60 keV have the
potential to give a reconstructed spectrum comparable with \textit{Fermi} GBM.
\bcol{Once the spectral response is
known, it can be incorporated into the spectral correction of
Eqn.~\eqref{eqn:corrspec}.}
Calibration uncertainties in the spectral response will affect source localisation. An associated
fact is that the spectral response of detector units will have a weak
dependence on source position. Although rigorous techniques exist, which  
solve for the source spectrum and position simultaneously~\citep{Balrog}, 
a simpler iterative approach for calibrating both the spectral response and 
the derived localisation is also possible.

These systematic uncertainties caused by imperfections in detector calibration, 
will be estimated post detector assembly by benchmarking 
simulations against calibration measurements as done in
\citet{validate,gapbrst}. 

\subsubsection{Background}
Background count rates during a GRB should be constant given the relatively short
duration of each GRB. Thus measurement of the background rates before and after
the GRB event enables its modeling and subtraction and the localisation
estimates \bcol{($\hat{\delta},\hat{\alpha}$)} will remain unaffected. 
The background measurement though, will have measurement errors. 
This will propagate through the routines to increase uncertainty on source
position. The background uncertainty is incorporated into each routine by
simulating the expected background as given
in \citet{fxie18}. \bcol{The expected background rate for localisation events
(single-hits above 50 keV) is $\sim760$ cts/s across the entire detector array.}

If $b_T$ is the total background counts measured in total time $t_b$ and $t_s$ is the
time duration of the GRB, then the expected number of background counts $b_m$ during
the GRB is simply
\begin{equation}
	b_m = b_T \cdot \frac{t_s}{t_b} \equiv r\cdot{b_T}
	\label{eqn:bgcts}
\end{equation}
Generally $r$ is less than unity as the window $t_b$ is made large enough to minimise uncertainty and measure a stable background rate. 
Let $\bcol{n = s + b}$ are the total detected counts, where $\bcol{s}$
gives the source (GRB) counts and $b$ the background counts. 
For the modulation curve routine, this simply increases the
uncertainty on counts in each outermost detector unit. 
The new uncertainties on each angular bin ($\phi_d$) will be 
\[ \sqrt{\bcol{s_i} + b_i(1+ r)}, \]
(instead of $\sqrt{\bcol{s_i}}$), where $b_i$ is 
the expected background  ($b_m$) in each detector unit $i$. The associated
increase in \bcol{$\sigma_\alpha,\sigma_\delta$} is found by fitting Eqn.~\eqref{eqn:modln} to
this data with higher uncertainties. 

The $\chi^2$ expression of Eqn.~\eqref{eqn:chimod} can be modified to similarly
account for the background uncertainty as 

\begin{equation}
	\chi^2(x,y) =  \sum_i \frac{(c_i - c_{tot} \cdot \bcol{\bar{m}_i(x,y)})^2}{c_{tot}
	\cdot \bcol{\bar{m}_i(x,y)} + b_i( 1 + r)}
	\label{eqn:chibg}
\end{equation}

The likelihood expression can also be modified to give Eqn.~\eqref{eqn:bayesbg},
where $\bcol{n_{mi}} =  c_{tot} \cdot \bcol{\bar{m}}_i(x,y) + b_i$ is  and $b_{Ti}$ is the measured
background (in time $t_b$) in detector unit $i$.

\begin{equation}
\begin{aligned}
	p(x,y | c_i) \propto \prod_i \left\{ \frac{\exp{-(\bcol{n_{mi}} + b_{Ti})}}{r} 
		\sum_{j=0}^{\infty} \left(\frac{\bcol{n_{mi}}^{\bcol{s_i} + j}\,
		b_{Ti}^{j/r}}{(\bcol{s_i} + j) ! (j/r) !}\right) \right\} \\
	\cdot \left(\sqrt{\frac{x^2 + y^2}{R^2}}\right)
	\label{eqn:bayesbg}
\end{aligned}
\end{equation}

\begin{figure}[!h]
	\centering
	\includegraphics[width=0.5\textwidth]{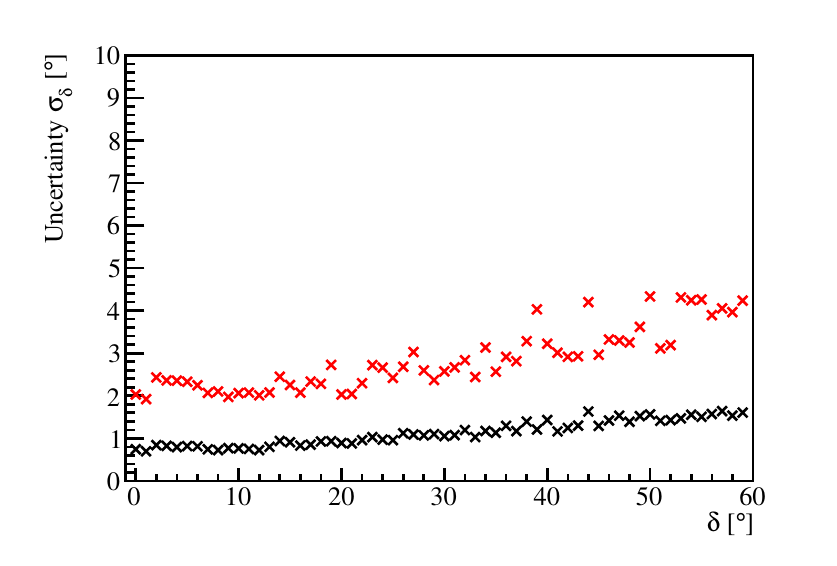}
	\caption{Increase in uncertainty due to background uncertainties (measured
		for 300s) in the $\chi^2$ routine for a median fluence GRB lasting 23s . Black points are before background correction and red
	points are after background correction. }
	\label{fig:bgcorr}
\end{figure}

The result of implementing these corrections is shown in Fig.~\ref{fig:bgcorr} for the $\chi^2$ routine. As seen from
the figure, background measurement uncertainties (for $t_b$ = 300s, $t_s$ = 20s) increase the localisation
uncertainty by $\sim 2^\circ$ (for a GRB with median fluence). These uncertainties
are expected to reduce as the orbital background measurement improves ($t_b$
increases). \bcol{This treatment assumes a constant background 
during the time window of measurement. The background rates
may vary during a GRB event. This can be treated by interpolating between the
measured rates before and after the GRB. In such cases, the
uncertainty on the background will be derived from the uncertainty of the
interpolation parameters, which will be Gaussian (and not Poisson) distributed.
The localisation uncertainty in such a case can be handled by re-writing the
likelihood term as done in case of PGSTAT in Xspec~\footnote{\url{https://heasarc.gsfc.nasa.gov/xanadu/xspec/manual/XSappendixStatistics.html}}.}

\subsubsection{ Use of the database}
\begin{figure}[!ht]
	\centering
	\includegraphics[width=0.5\textwidth]{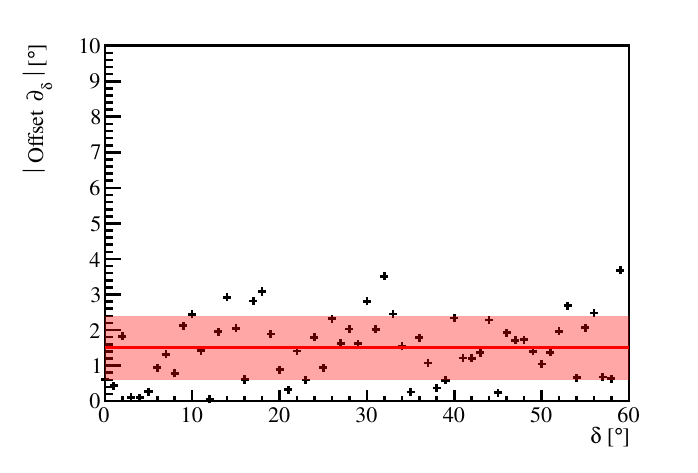}
	\caption{Systematic offsets \bcol{($\partial_\delta$)} in the 
	$\chi^2$ routine. The mean and $1\sigma$ deviation for offsets are highlighted.}
	\label{fig:offs}
\end{figure}

	Some of the assumptions used to create the database give rise to systematic offsets
	(measured as the angular distances \bcol{$\partial_\delta,\partial_\alpha$} between the actual
	and estimated source positions). This is shown in Fig.~\ref{fig:offs} for
	the $\chi^2$ database routines. The simulated
	database is made with a fluence of 200~ph/cm$^2$ and lowering this to 20
	ph/cm$^2$ increases the average offsets from 
	$1.5^\circ$ to $3.5^\circ$. 
	This shows that the use of Monte-Carlo techniques involved 
	in creating the database introduces an additional source of uncertainty due
	to Poisson fluctuations in the database counts. As average offsets of
	$\sim1.5^\circ$ do not affect polarisation measurements adversely (see Fig.~\ref{fig:vic}), the
	database (with 200~ph/cm$^2$) is used by default.  

\begin{figure}[!h]
	\centering
	\includegraphics{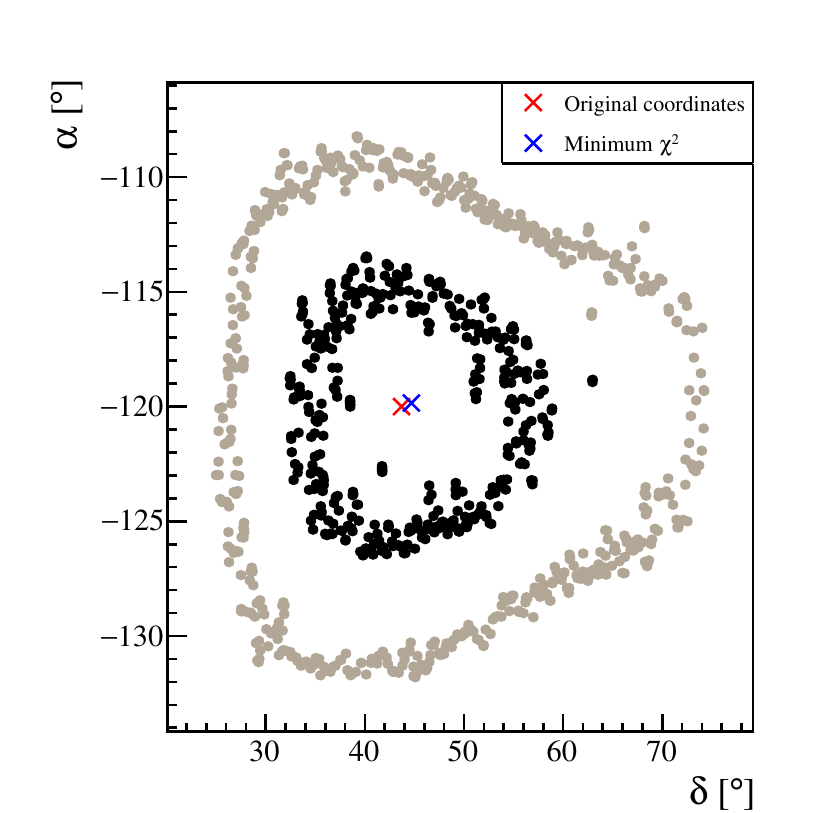}
	\caption{Uncertainty contours (1$\sigma$ and 2$\sigma$) using the $\chi^2$
	routine for a weak fluence GRB (2 ph/cm$^2$). }
	\label{fig:chicont}
\end{figure}

	\bcol{For the routines using the database, interpolation enables estimation of uncertainty
	in between the non-regularly spaced grid points on the sky. Contour-based non-interpolative
	methods can only be used for GRBs with low fluence as in such
	cases (see next section) the uncertainty is much greater than the grid-size,
	thereby allowing construction of smooth contours. This is seen in
	Fig.~\ref{fig:chicont} for a GRB with very weak fluence.} 

	The $\chi^2$ routine though, uses \bcol{parabolic interpolation and}
	standard error propagation to obtain uncertainty in sky co-ordinates.
	This causes the obtained localisation uncertainty to have a systematic azimuthal dependence (as seen in
	Fig.~\ref{fig:errad}). The amplitude of this variation ($<1^\circ$) is much
	smaller than other systematic effects. The likelihood routine needs a smaller
	grid spacing to \bcol{obtain} posterior distributions and uncertainty estimates. One
	solution is to make an additional database in a small
	region of the sky (say $5^\circ\times5^\circ$) with a finer grid size and
	increased fluence as discussed in Section~\ref{ss:compar}.

\subsubsection{ GRB fluence effects}
\begin{figure}[!ht]
	\centering
	\includegraphics[width=0.5\textwidth]{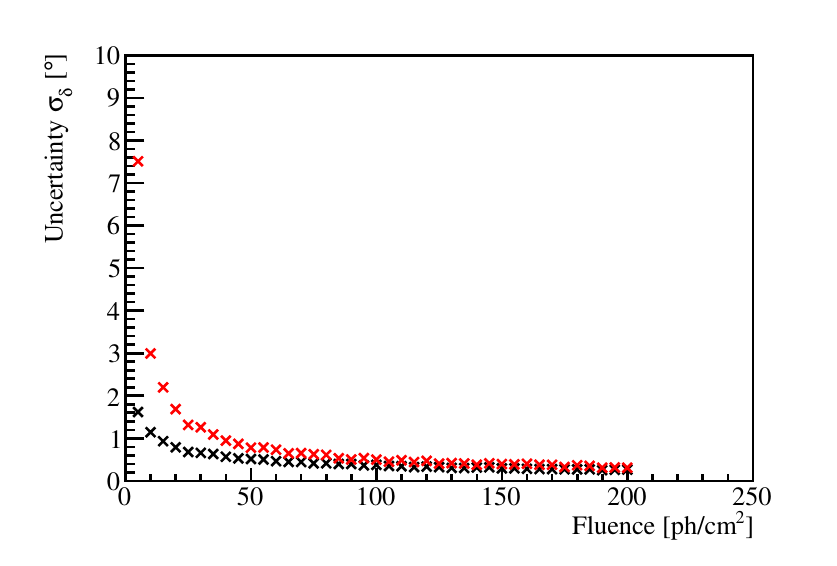}
	\caption{Uncertainty \bcol{$\sigma_{\delta}$} for the $\chi^2$ routine as a
		function of source fluence with (red) and without (black) background uncertainty
($t_b =$ 300s, $t_s =$ 20s). }
	\label{fig:resflnc}
\end{figure}

	Photon counting statistics ensure that bright GRBs
	are better localised than faint GRBs. This is seen in
	Fig.~\ref{fig:resflnc} and is true irrespective of the routine used. 
	SPHiNX can detect polarisation for GRBs with fluence $\gtrsim
	20$~ph/cm$^2$~\citep{pearce18}. For these GRBs, the localisation
	uncertainty in \bcol{$\delta$} using $\chi^2$ routine can be computed as 
	\begin{equation}
		\sigma = \sqrt{\bcol{\sigma_\delta^2 + \partial_{off}^2} + \sigma_{cal}^2}, 
	\label{eqn:totuncer}
	\end{equation}
	where $\bcol{\sigma_\delta} < 2.2^\circ$ is uncertainty including background and
	$\bcol{\partial_{off}}\sim1.5^\circ$ is the average offset. If we take uncertainty
	due to calibration systematics to be $\sigma_{cal}\sim2^\circ$, then for these
	GRBs $\sigma < 3.4^\circ$ and the
	MDP does not increase significantly ($<5\%$) for all GRBs within the FOV
	(Fig.~\ref{fig:vic}). SPHiNX can achieve a reasonable localisation
	($\bcol{\sigma_\delta}<5^\circ$) for weaker GRBs too (with fluence $>$10~ph/cm$^2$). 

\subsection{Comparison of routines}\label{ss:compar}
The three routines considered in this paper are compared based on the
assumptions, obtained uncertainty estimates, computational resources needed and
possible use cases in Table~\ref{tab:comp}. The computations were performed 
on one of the six cores of an Intel Xeon E5 CPU (assisted by 16 GBs of RAM space)
and do not include time required for background correction. The
database was stored on an external 3 Terabyte magnetic hard-disk connected via USB (which
increases read times required to access the database). 

\begin{landscape}
\begin{table}
	\centering
	\caption{Comparison of Inversion routines}
	\label{tab:comp}
	\begin{tabular}{|p{3cm}|p{5.5cm}|p{5.5cm}|p{5.5cm}|}
\hline
 & Routine-I (Modulation) & Routine-II ($\chi^2$) & Routine-III (Likelihood) \\
\hline
Approximations & $\bullet$ Uses subset of detected counts. &
				 $\bullet$ Uses all available counts. &
				 $\bullet$ Uses all available counts. 
				 \\
				 &  $\bullet$ Empirical expression as mapping function. &
				 $\bullet$ Simulated database as mapping function. &
				 $\bullet$ Simulated database as mapping function. 
				 \\
			  &	 &
				 $\bullet$ Gaussian approximation to Poisson distributed counts. &
				 $\bullet$ Poisson distribution directly used.
				 \\
			  &	 &
				 $\bullet$ Uncertainty estimate uses error propagation (assuming independent $\sigma_x,\sigma_y$)
				 and parabola fit around $\chi^2_{min}$. &
				 $\bullet$ Uncertainty estimate uses linear interpolation and dependent on grid size of database.
				 \\
				 \hline

 Resources	  &	 $\bullet$ Fast one-step execution ($\sim$~0.35 seconds). &
				 $\bullet$ Slow execution (5 hours - spectral correction ; 
				 35 sec - database read, minima search and uncertainty
			 computation on standard grid). &
			 	$\bullet$ Slow execution (7 hours - creation of a fine-grid
					small region database, 5 hours - spectral correction ; 
					5 sec - database read, maxima search and uncertainty
				estimation on smaller database). 
				 \\
			&   $\bullet$ No need for a database. Additional storage and read / write to hard-disk not required. &
				$\bullet$ Additional storage for an external database necessary. &
				$\bullet$ Additional finely gridded database ($\sim 7$ Gigabytes) necessary for
				obtaining uncertainty.
				\\
				\hline

				Uncertainty &	$\bullet$ Large localisation uncertainty.
				$\bcol{\sigma_\delta}~\simeq~4^\circ$ without offset and background correction. 
				&
				$\bullet$ Low uncertainty. $\bcol{\sigma_\delta} \simeq 1^\circ$ without offset and background correction.
				&
				$\bullet$ Low uncertainty. $\bcol{\sigma_\delta} \sim 0.2^\circ$ without offset and background correction.
				\\
				\hline
Possible use &	$\bullet$ Quick estimates for light-weight on-board computation. \newline
				$\bullet$ First level estimate used to reduce and refine search space for a fine grid Likelihood based localisation.
				&
				$\bullet$ Provides sufficiently accurate localisation to use with polarisation analysis.
				&
				$\bullet$ Can be combined with Routine-I for fine grid search
				over a small region to obtain very accurate localisation. Allows
				incorporation of a priori information.
				\\
				\hline
				
\end{tabular}
\end{table}
\end{landscape}

\section{Conclusion}\label{sec:conclusion}
This paper discusses three GRB localisation routines for use with the SPHiNX mission
concept. On-board computation of localisation parameters is not foreseen in the
current mission design due to
the limited downlink cadence (of one downlink per day) which makes real-time
localisation unfeasible. 
The main use of localisation will be to
determine the source position offline for use with subsequent polarisation analysis.

Polarisation properties will not be significantly affected for a \bcol{zenith}
angle uncertainty of $\bcol{\sigma_\delta} \lesssim 5^\circ$. For median fluence GRBs, the localisation uncertainty is
mainly driven by the presence of systematic offsets and uncertainty in
background measurements. For higher fluence GRBs, these
uncertainties reduce considerably. Localisation will not be undertaken for
GRBs with weaker fluence. 

The modulation curve routine, with its light-weight 
implementation and independence from the GRB spectrum is suitable for quick
computations. This routine has a higher localisation uncertainty as it does
not utilise information from all detector units. However, it can be used to get a rough estimate of the
source position which can in turn be used to make a finely gridded database over
a small region. This finer database can then be used with the likelihood routine
to obtain a more accurate localisation. This coarse but fast estimate can be used to issue GCN notices in future GRB mission
proposals which have higher downlink cadence provided the mission design
preserves the azimuthal symmetry needed for generating such modulation curves.

The $\chi^2$ routine is ideally suited for offline use with simple approximations to compute the uncertainty. 
This routine gives an estimate within the required accuracy for polarisation
analysis. While better techniques (like the contour-based uncertainty estimate)
can be used to improve accuracy, the likelihood based routine is preferred in
such cases as it makes fewer assumptions. 

The likelihood based routine makes the highest demand on use of computational
resources (especially in the nested sum computation needed for background
correction). It also needs a database with a finer grid-size. 
However, it provides more accurate positions (as it uses fewer assumptions). If
source position accuracy is prioritised, then this method can be coupled to a database with higher
fluence and smaller grid size (around the coarsely localised modulation routine
estimate) to get an accurate position estimate. Such requirements may arise if 
SPHiNX localisations are needed for detailed follow up studies of a particular GRB. 

More sophisticated routines based on neural networks or maximising other statistics (like the correlation between detected count map and database count map) can be 
also be used for localisation. It is unlikely though, that the accuracy will be
significantly better than for the likelihood routine as the position accuracy is
mainly limited by photon statistics and design / placement of the detectors.
In general, localisation for Compton polarimeters can be improved if the
placement and shielding of detector units can
be carefully optimised (without compromising polarisation sensitivity) by using studies similar to the ones done for coded aperture
masks~\citep{fenican}. 

\section*{Acknowledgments}
 \addcontentsline{toc}{section}{Acknowledgments}
The authors thank Victor Mikhalev for providing Fig.~\ref{fig:vic} and for useful discussions on statistics and uncertainty
estimations. Funding received from the Swedish National Space Agency (grant
number 232/16) is gratefully acknowledged. 


\bibliographystyle{mnras}   
\bibliography{loc}   

\section*{About the authors}
\addcontentsline{toc}{section}{About the authors}
\vspace{2ex}\noindent\textbf{Lea Heckmann} recently joined the 
MAGIC group at the Max Planck Institute for Physics, Munich as a PhD student.
She received her MSc degree in engineering physics from the Vienna University of
Technology in 2018. The main part of her master thesis was conducted at 
KTH Royal Institute of Technology in Stockholm,
focusing on simulation studies and data analysis for the SPHiNX mission.

\vspace{2ex}\noindent\textbf{Nirmal Iyer} is currently a post-doctoral
researcher working with the SPHiNX and X-Calibur projects at KTH Royal 
Institute of Technology, Stockholm. He received his PhD
from Indian Institute of Science, Bangalore in 2016 for work with design
and development of an X-ray sky monitor. He is interested in
design and optimisation of X-ray detectors for astronomy. 

\vspace{2ex}\noindent\textbf{M\'ozsi Kiss} is working as a Researcher in Astroparticle Physics at KTH Royal
Institute of Technology in Stockholm since 2011. He conducted his Master's
thesis work at the Stanford Linear Accelerator Center, US, working on
Compton-based polarimetry in 2006, and received his Ph.D. from KTH Royal
Institute of Technology in 2011 working on the construction and testing of the
PoGO/PoGOLite balloon-borne Compton polarimeter. Current interests include
detector construction, scientific ballooning, satellite-based payloads and
X-ray/gamma-ray polarimetry.

\vspace{2ex}\noindent\textbf{Mark Pearce} is professor of physics at KTH Royal Institute of
Technology in Stockholm since 2007. He received a Ph.D. in 1996 from the University of
Birmingham, UK, for research in experimental particle physics. He currently
focusses on instrument development for X-/gamma-ray astrophysics and the study
of compact astrophysical objects using X-ray polarimetry. He is Principal
Investigator of the PoGO+ and SPHiNX missions for hard X-ray polarimetry 
and recently joined a related mission, X-Calibur. Since
2012, he is Head of the KTH Physics Department.

\vspace{2ex}\noindent\textbf{Fei Xie} is a postdoc in INAF, Osservatorio
Astronomico di Cagliari. She received her Ph.D. in astroparticle physics
from the Institute of High Energy Physics, University of Chinese Academy of
Sciences in 2016. After that, she joined the particle and astroparticle
physics group in KTH Royal Institute of Technology and mainly focused on
X-ray polarimetry, including background simulation, instrument optimisation.
From the second half of 2018, she began to work on IXPE (Imaging X-ray
Polarimetry Explorer) project in Italy.

\end{document}